\documentclass[%
 prb,
 amsmath,amssymb,floatfix,
 reprint,%
]{revtex4-1}

\usepackage{graphicx}
\usepackage{dcolumn}
\usepackage{bm}
\usepackage{amsmath}
\usepackage[utf8]{inputenc}
\usepackage[T1]{fontenc}
\usepackage{etoolbox}
\usepackage{color}

\newcommand{\up}{{\uparrow}}
\newcommand{\down}{{\downarrow}}
\newcommand{\hc}{\text{h.c.}}

\newcommand{\img}{\mathrm{i}}

\newcommand{\Lleg}{\mathrm{L}}
\newcommand{\Rleg}{\mathrm{R}}

\begin{document}

\title{A simple electronic ladder model harboring $\mathbb{Z}_4$ parafermions}
\author{Botond Osv\'ath}
  \affiliation{Department of Physics of Complex Systems, ELTE E\"otv\"os Lor\'and University, H 1117, Budapest, Hungary}
\author{Gergely Barcza}
  \affiliation{Strongly Correlated Systems Lend\"{u}let "Momentum" Research Group Wigner Research Centre for Physics, H-1525, Budapest, Hungary}
  \affiliation{MTA–ELTE Lend\"{u}let "Momentum" NewQubit Research Group, Pázmány Péter, Sétány 1/A, 1117 Budapest, Hungary}
\author{Örs Legeza}
  \affiliation{Strongly Correlated Systems Lend\"{u}let "Momentum" Research Group Wigner Research Centre for Physics, H-1525, Budapest, Hungary}
  \affiliation{Institute for Advanced Study, Technical University of Munich, Lichtenbergstrasse 2a, 85748 Garching, Germany}
\author{Balázs Dóra}
  \affiliation{Department of Theoretical Physics, Institute of Physics, Budapest University of Technology and Economics, M\H uegyetem rkp. 3., H-1111 Budapest, Hungary}
\author{L\'aszl\'o Oroszl\'any}
  \affiliation{Department of Physics of Complex Systems, ELTE E\"otv\"os Lor\'and University, H 1117, Budapest, Hungary}
  \affiliation{Wigner Research Centre for Physics, H-1525, Budapest, Hungary}
  \email{laszlo.oroszlany@ttk.elte.hu}

\date{\today}

\begin{abstract}
Parafermions are anyons with the potential for realizing non-local qubits that are resilient to local perturbations. 
Compared to Majorana zero modes, braiding of parafermions implements an extended set of topologically protected quantum gates. 
This, however, comes at the price that parafermionic zero modes can not be realized in the absence of strong interactions
posing a challenge for their theoretical depiction. 
In the present work, we construct a simple lattice model for interacting spinful electrons with parafermionic zero energy modes.
The explicit microscopic nature of the considered model highlights new realization avenues for these exotic excitations in recently fabricated quantum dot arrays.
By density matrix renormalization group calculations, we identify a broad range of parameters, with well-localized zero modes, whose parafermionic nature is substantiated by their unique $8\pi$ periodic Josephson spectrum. 
\end{abstract}

\maketitle



\section{Introduction}

As outlined at the turn of the century, key requirements for the realization of a quantum computer are qubits with long decoherence times and a universal set of robust quantum gates acting on them \cite{Divincenzo_2000}.
Topological quantum computers, where quantum information is encoded in non-local quasiparticles, promise to address these issues at the hardware level \cite{Das_Sarma_RevModPhys.80.1083,Aasen_Majorana_PhysRevX.6.031016}.
Majorana fermions \cite{Kitaev_2001} and their fractionalized counterparts, parafermions \cite{fendley_parafermion}, are such excitations in topological superconductors \cite{alicea_quantum_comp,alicea_fendley}.
Parafermions are classified by a $\mathbb{Z}_d$ index with $d=2$ corresponding to Majorana fermions. The characteristic non-Abelian exchange statistics of parafermionic excitations can be exploited to implement topologically protected quantum gates. Braiding of $d>2$ parafermion zero modes implements an extended set of quantum operations compared to Majorana zero modes~\cite{hutter2016quantum}.

Blueprints for realizing parafermions rely on electron-electron interactions \cite{alicea_fendley}. Furthermore, some of these proposals require combining superconductivity and fractional quantum Hall edge modes, whose realization is challenging as the magnetic field needed for stabilizing the quantum Hall phase is detrimental for conventional superconductors \cite{FQH_para_clarke2013exotic,FQH_para_PhysRevX.2.041002}.
Several recent theoretical works have proposed platforms for realizing $\mathbb{Z}_4$ parafermion zero modes circumventing this conundrum. These approaches are based on strongly interacting quantum spin Hall (QSH) insulators coupled to superconductors and do not require a substantial external magnetic field \cite{zhang,orth2015non}. In these proposals, interactions open a gap in the edge states while preserving time-reversal symmetry. Parafermion modes emerge at interfaces between edge regions gapped by interactions and proximity-coupled superconducting sections. The $\mathbb{Z}_4$ parafermionic modes manifest a fourfold degenerate ground state and are characterized by $8\pi$-periodic fractional Josephson effect.
Interestingly, this effect can be realized without an extended region of strong interactions. In recent proposals, it was suggested that coupling an impurity quantum spin via an anisotropic exchange term to the Josephson junction formed at a QSH edge can exhibit $8\pi$-periodicity \cite{Peng2016,Vinkler2017}.
Weak interactions in a constriction of topological insulators can also stabilize $8\pi$ Josephson signal \cite{Fleckenstein2019}. The existence of parafermionic zero modes was also investigated in spinless \cite{Klinovaja_PhysRevLett.112.246403} and spinful \cite{Klinovaja_PhysRevB.90.045118} nano-wire setups.

All the studies above regarding the physical implementation of parafermionic phases rely on the bosonization technique, thus neglecting high-energy or lattice-scale effects\cite{giamarchi}. 
Alternatively, lattice models can be used to study the rich landscape defined by parafermionic excitations.
Density matrix renormalization group (DMRG) method \cite{white_dmrg_1} has been applied to 
investigate dynamical excitations in parafermionic chains \cite{Lado_PhysRevResearch.3.013095}.
There is a natural mapping between $\mathbb{Z}_4$ parafermionic and spinful fermionic degrees of freedom since they both encode a four-dimensional local Hilbert space. The algebraic structure of this mapping has recently been thoroughly explored \cite{Fermionized_parafermions_PhysRevB.98.085143}. 
The fermionized version of parafermion chains has also been investigated, yielding an electronic lattice model with exotic components, such as three-particle interactions as well as occupation and spin-dependent hopping \cite{calzona2018z,Teixeira2021}.


Here, we follow a more realistic approach inspired by experimentally feasible proposals utilizing the helical QSH edge modes. In Sect.~\ref{sect:model}, we construct a lattice model for spinful electrons with the potential to host $\mathbb{Z}_4$ parafermionic zero modes. 
Utilizing the DMRG approach we explore its rich phase diagram and demonstrate the presence of the parafermionic states in Sects.~\ref{sect:phases} and \ref{sect:para} respectively.
In Sect.~\ref{sect:Heisenberg} we discuss the impact of isotropic exchange interaction on the stability of the parafermionic phase. As summarized in Sect.~\ref{sect:discuss}, the simplicity of our model highlights new realization avenues of parafermionic zero modes in highly tunable quantum dot arrays.
Further details of our calculations and analysis are given in the appendix.

\section{The model}
\label{sect:model}
\begin{figure}[!t]
  \includegraphics[width=0.45\textwidth]{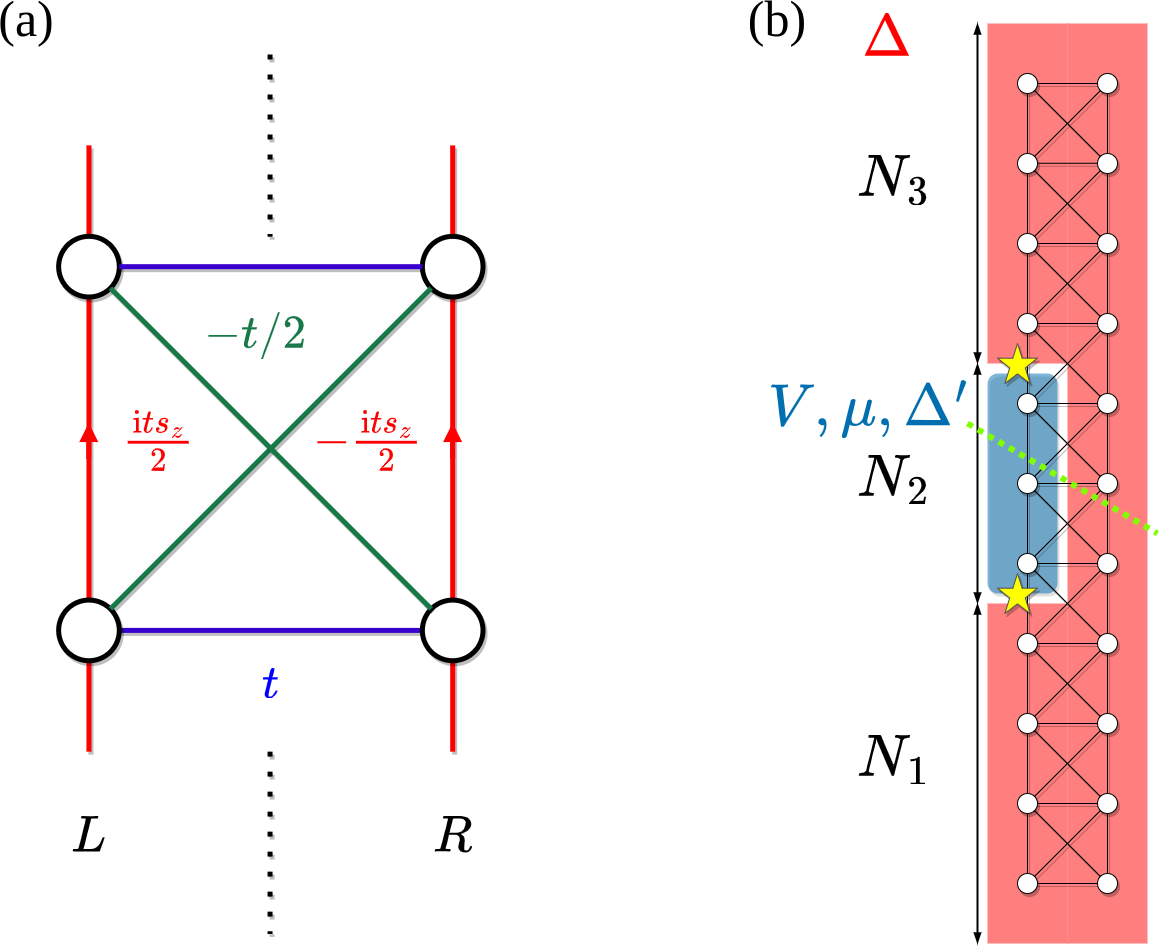}
  \caption{(a) Sketch of the kinetic term, detailing all single particle hopping processes along the ladder. Note the opposite sign of the complex amplitudes of the parallel hopping processes along the two legs of the ladder. These refer to the hopping direction marked by the arrows.   
  (b) Configuration of a finite size system used for obtaining the phase diagram and excitation spectrum. Red shade denotes the region of superconductivity with pair potential strength $\Delta$. Blue color signals a region with interactions of strength $V$, homogeneous potential $\mu$, and superconducting term with pair potential $\Delta'$. The green dashed cut denotes the interface at which the entanglement entropy is obtained for the phase diagram calculations.}
  \label{fig:sketch_kinetic}
\end{figure}

In the following, we introduce a ladder Hamiltonian acting on spinful electrons which captures the essential properties of the edge states of two-dimensional topological insulators without explicitly treating the insulating bulk. We build on previous works where a similar lattice model was applied for modelling free fermions~\cite{Creutz_PhysRevLett.83.2636,Creutz_Hetenyi_gholizadeh2018extended,Creutz_Dora_PhysRevB.88.205401}, helical Majorana modes subject to interactions~\cite{lattice_model_grover2014emergent,lattice_model_PhysRevB.102.165123}, and the edge states of fractional topological insulators\cite{Santos_Beri_PhysRevB.100.235122}.
In the proposed model each electronic site has local spin degrees of freedom denoted by $\up$ and $\down$ while the left and right leg of the ladder will be referred to simply by the labels $\Lleg$ and $\Rleg$ respectively. 
We write the Hamiltonian of the system as the sum of three parts:
\begin{equation}
H=H_\text{kin}+H_\text{sc}+H_\text{int}\,.
\label{eq:Ham}
\end{equation}
The first term  describes the kinetic contributions, capturing propagation along the legs and hopping across the rungs of the ladder,
\begin{align}\label{eq:kin}
{\displaystyle H}_{\text{kin}} & =\sum_{m} c_{m}^{\dagger} \left( 
 -\mu_m s_0 \otimes \zeta_0 + t s_0 \otimes \zeta_x
\right)c_{m} \\
 & -\frac{t}{2}\sum_{m} c_{m+1}^{\dagger} \left( i
 s_z \otimes \zeta_z +  s_0 \otimes \zeta_x
\right)c_{m} + \hc \nonumber \,.
\end{align}
Here, $s_\alpha$ and $\zeta_\alpha$ are Pauli matrices acting on the spin and leg degrees of freedom respectively, and 
$ c^\dagger_{m} = \left ( c^\dagger_{m, \Lleg, \up},c^\dagger_{m, \Rleg, \up},c^\dagger_{m, \Lleg, \down},c^\dagger_{m, \Rleg, \down} \right)$ where $c^\dagger_{m,\zeta,s}$ 
denotes the creation operator of an electron with spin projection $s \in\{\up,\down\}$ on-site $m$ of leg $\zeta\in\{\Lleg,\Rleg\}$. $t$ serves as an overall energy scale for the system, while $\mu_m$ is a site-dependent potential. 
A visual representation of the different considered hopping processes is shown in Fig.~\ref{fig:sketch_kinetic}~(a).

For low energies, the kinetic term \eqref{eq:kin} describes the propagation of helical particles which can be made more explicit if we take the Fourier transform of \eqref{eq:kin} along the ladder, by introducing operators $c_k=\sum_m \mathrm{e}^{\img mk}c_m$. Thus the kinetic term, for uniform values of the chemical potential $\mu_m=\mu$, becomes $H_\text{kin}=\sum_k c^\dagger_k \mathcal{H}(k) c_k$. For low energies, this term can be analyzed by the envelope function approximation around $k=0$ yielding  
\begin{equation}
\mathcal{H}(k) \approx -\mu s_0 \otimes \zeta_0 + k t s_z\otimes \zeta_z \,.
\label{eq:EFA}    
\end{equation}
Crucially no terms are mixing the two legs, thus for low-energy helical particles, the two legs are decoupled just as they would be for two spatially separated edges of a large two-dimensional topological insulator.  We note that our model can also be employed for investigating disjoint edges of two distinct topological insulators coupled by a Kondo impurity\cite{Biswas_Kondo_PhysRevB.109.155119}. 

The second term in the Hamiltonian \eqref{eq:Ham} describes proximity to an $s$-wave superconductor,
\begin{equation}
H_{\text{sc}}=\sum_{m,\zeta}\Delta_{m,\zeta}\left[c_{m,\zeta,\up}^{\dagger}c_{m,\zeta,\down}^{\dagger}+\hc\right],
\label{eq:superconductivity}
\end{equation} with a site and leg-dependent pair potential $\Delta_{m,\zeta}$.

The last term, $H_{\text{int}}$, describes a short ranged microscopic interaction  
\begin{equation}
    H_{\text{int}}=  \sum_{m,\zeta}V_{m,\zeta}\left[c_{m,\zeta,\up}^{\dagger}c_{m,\zeta,\down}c_{m+1,\zeta,\up}^{\dagger}c_{m+1,\zeta,\down}+\hc \right]\,.
    \label{eq:interaction}
\end{equation}
Introducing 
$c^\dagger_{m,\zeta} = \left ( c^\dagger_{m, \zeta, \up},c^\dagger_{m, \zeta, \down} \right )$
and rewriting the interaction term with operators $S^\alpha_{m,\zeta}=c^\dagger_{m,\zeta}s_\alpha c_{m,\zeta}$ we get
\begin{equation}
    H_{\text{int}}=  \sum_{m,\zeta}\frac{V_{m,\zeta}}{2}\left[ S^x_{m,\zeta}S^x_{m+1,\zeta}-S^y_{m,\zeta}S^y_{m+1,\zeta}\right],
    \label{eq:interaction-spin}
\end{equation}
that is, this term describes an anisotropic symmetric exchange coupling of strength $V_{m,\zeta}$ between electron spins on neighboring sites along a leg. 

An important aspect of this model is that it can be used to effectively circumvent fermion doubling \cite{montvay_munster_1994}, without breaking time-reversal symmetry albeit at the cost of dismissing charge conservation. For instance in a configuration depicted in Fig.~\ref{fig:sketch_kinetic} (b) a fixed, large value of $\Delta$ gaps the whole red region thus allowing the exploration of effects of arbitrary local interactions on the helical states localized in the blue region.
\begin{figure}[!t]
  \includegraphics[width=0.45\textwidth]{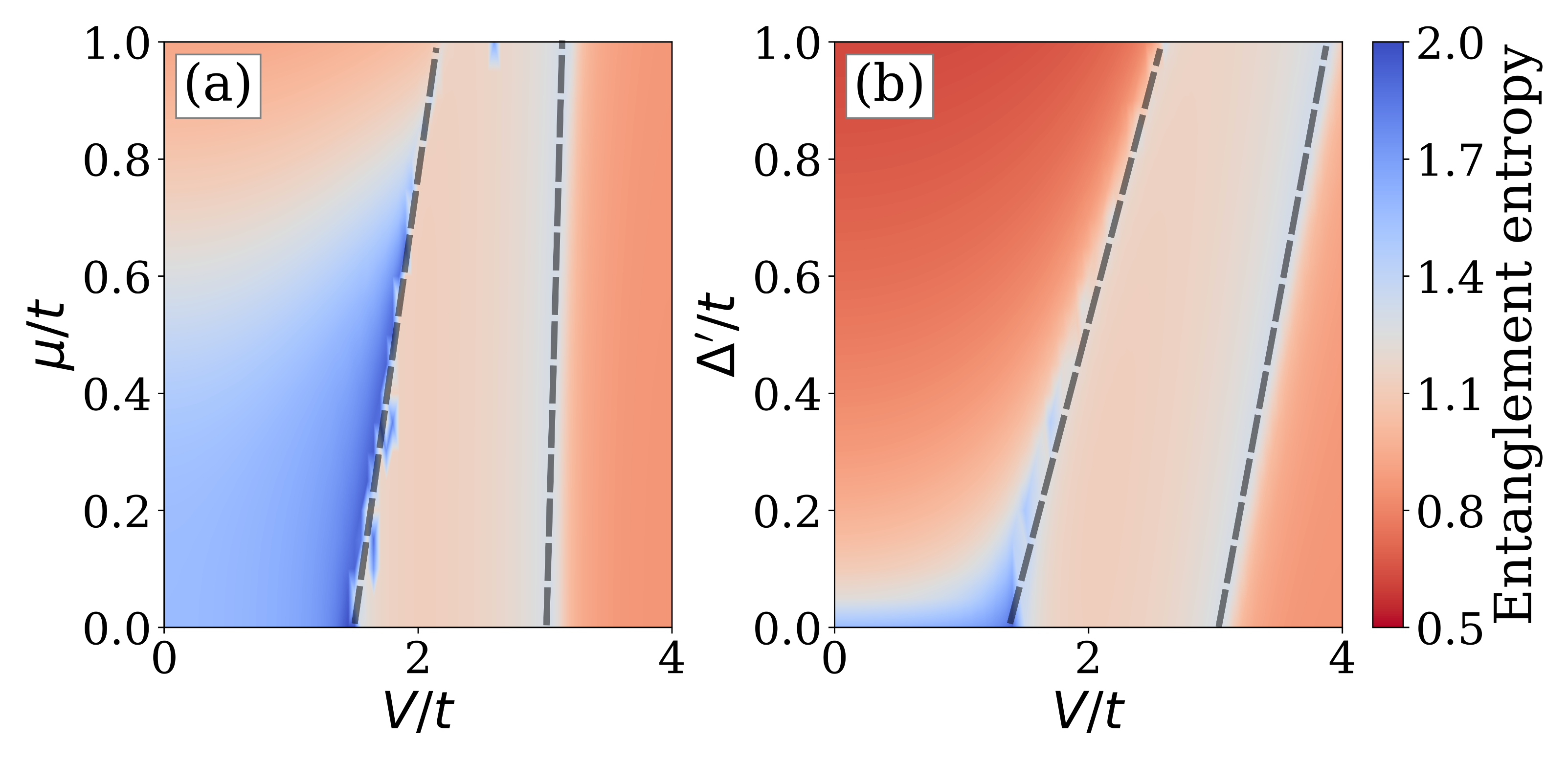}
  \caption{Entanglement entropy for model setup shown in Fig.~\ref{fig:sketch_kinetic}(b)  as a function of interaction strength $V$ for  $N_1=N_3=20$ and $N_2=100$ and $\Delta/t=1.0$. In subfigure (a) the chemical potential $\mu$, while in (b) the pair potential $\Delta'$ is varied. Dashed lines are a guide to the eye,  indicate phase boundaries. 
  }
  \label{fig:phases}
\end{figure}

\section{Phase diagram}
\label{sect:phases}
We explore the phase diagram of the proposed model \eqref{eq:Ham} with the DMRG~\cite{itensor,budapest_qcdmrg,SM} method as it is an ideal tool for characterizing quasi-one-dimensional systems. 
Quantum phase transitions can be detected by the anomaly of entanglement measures~\cite{Wootters1998,Osborne2002,Osterloh2002,Legeza2006}. In practice we computed the half system von Neumann entropy~\cite{Mund2009} for the geometry shown in Fig.~\ref{fig:sketch_kinetic} (b). {The technical details of the used DMRG calculations are presented in Appendix~\ref{sect:App_DMRG}, in this section we focus on the physical interpretation of the phase diagram.}

{To study the stability of the potential phases emerging by tuning the interaction strength, we investigate the phase diagram under general conditions controlled by parameters $\mu$ and $\Delta'$.}
The obtained entanglement entropy maps are depicted in Fig.~\ref{fig:phases}.
For $\Delta'=0$, three phases are clearly discernible in Fig.~\ref{fig:phases} (a). 
Allowing for finite $\Delta'$ shown in Fig.~\ref{fig:phases} (b), for small enough interactions any finite value of $\Delta'$ gaps the system thus pushing the middle region of the system where interactions are active (bluish part in Fig.~\ref{fig:sketch_kinetic} (b)) to the same superconducting phase as the rest. A hallmark of this is the marked drop in entanglement entropy as $\Delta'$ is increased. For large enough interaction strength, the two phases observed for $\Delta'/t=0$ remain stable even for large $\Delta'$. 

\begin{figure}[!t]
  \includegraphics[width=0.45\textwidth]{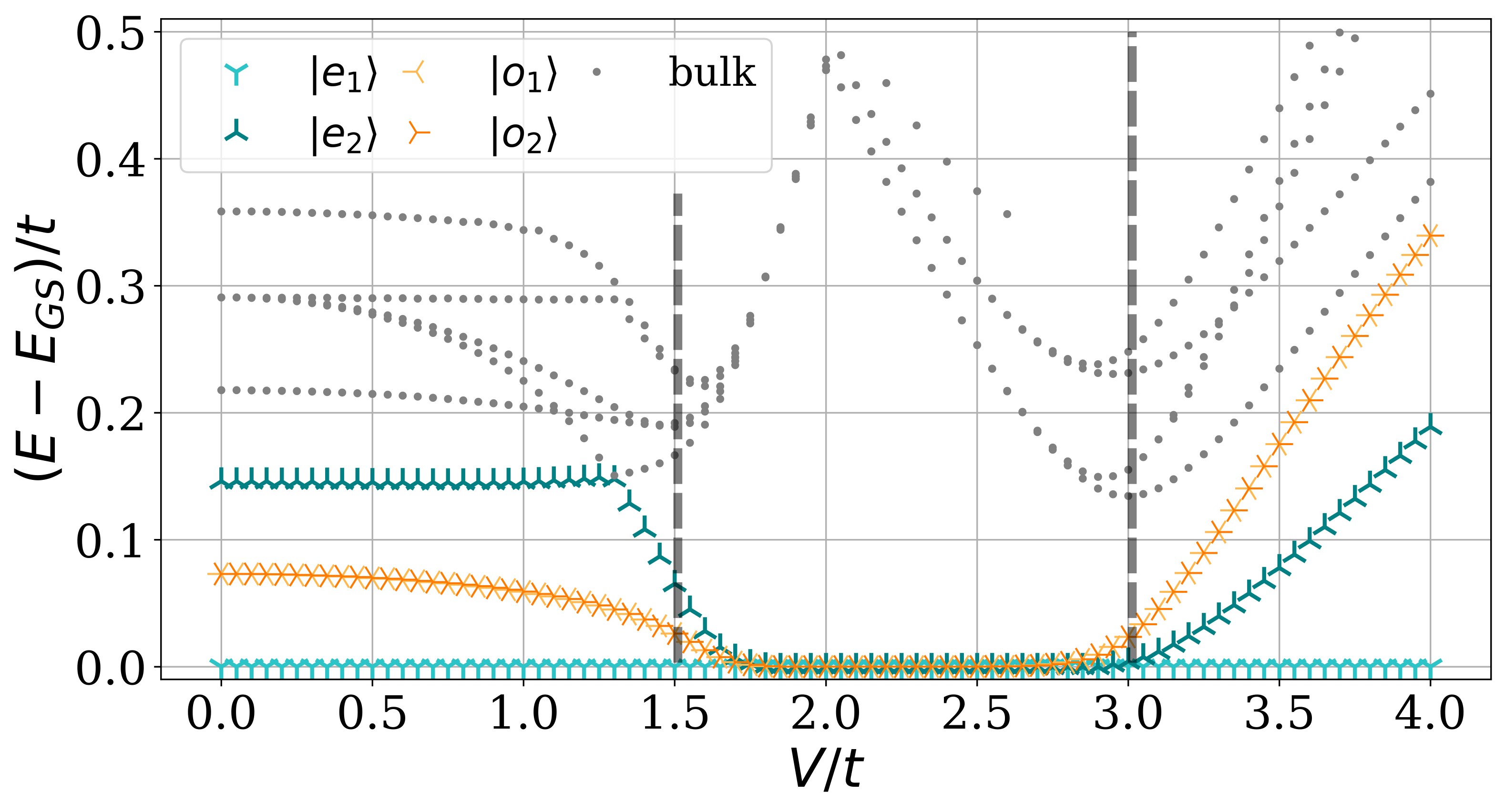}
  \caption{Excitation spectrum of the considered model as the function of the interaction strength $V$. For $N_i=20$ and $\mu/t=0$ in the configuration depicted in Fig.~\ref{fig:sketch_kinetic} (b). The two lowest energy even and odd parity states are marked by $|e_i\rangle$ and $|o_i\rangle$ with $i\in\{1,2\}$. Vertical dashed lines mark the same boundaries as in Fig.~\ref{fig:phases}}
  \label{fig:excitation_spectrum}
\end{figure}

Note that for large enough $\mu/V$ or $\Delta'/V$ the system reverts effectively to a phase that can be described by a non-interacting theory. For large $\mu$ the separation of the states to the two legs of the ladder does not hold, and the system will revert back to a metallic behaviour. For large $\Delta'$ the superconducting correlations overpower effects due to interactions, turning the whole system to a conventional superconducting phase.
To distinguish the observed phases, in Fig.~\ref{fig:excitation_spectrum} we now study the low-energy many-body excitation spectrum obtained by the DMRG approach~\cite{stoudenmire_excited_2012studying}  as the function of the interaction strength $V$ fixing $\Delta'/t=\mu/t=0$  for the sake of simplicity. Three different phases can be identified in complete agreement with our entropy analysis. Notably, for weak interaction, up to around $V/t=1.5$, the system has a well-defined ground state with even fermion parity, and the first excited state is a doubly degenerate odd state. In the thermodynamic limit, the spectrum of this phase shows a metallic character with a vanishing excitation gap as demonstrated in Appendix~\ref{sect:App_spectrum_scaling}. 
For intermediate interaction strengths, beyond the reach of perturbation theory, a phase with a fourfold degenerate ground state emerges with a considerable gap in the excitation spectrum. We label the states in this degenerate ground state manifold as $|e_i\rangle$ for even parity and $|o_i\rangle$ for odd parity with $i\in\{1,2\}$.
This spectral structure is found to be independent of the size of the system (see Fig.~\ref{fig:lengthscales} in  Appendix~\ref{sect:App_spectrum_scaling}). 
At around $V/t=3$ a second phase transition is observed. For stronger interaction strengths, the degeneracy is lifted again and the gap increases linearly with $V$. 

Note that the four zero-energy modes observed in our designed model at moderate interaction strengths have the potential to host  $\mathbb{Z}_4$ parafermions. 
In the following we show that in this phase the system is indeed characterized by parafermionic zero modes.

\begin{figure}[!b]
  \includegraphics[width=0.45\textwidth]{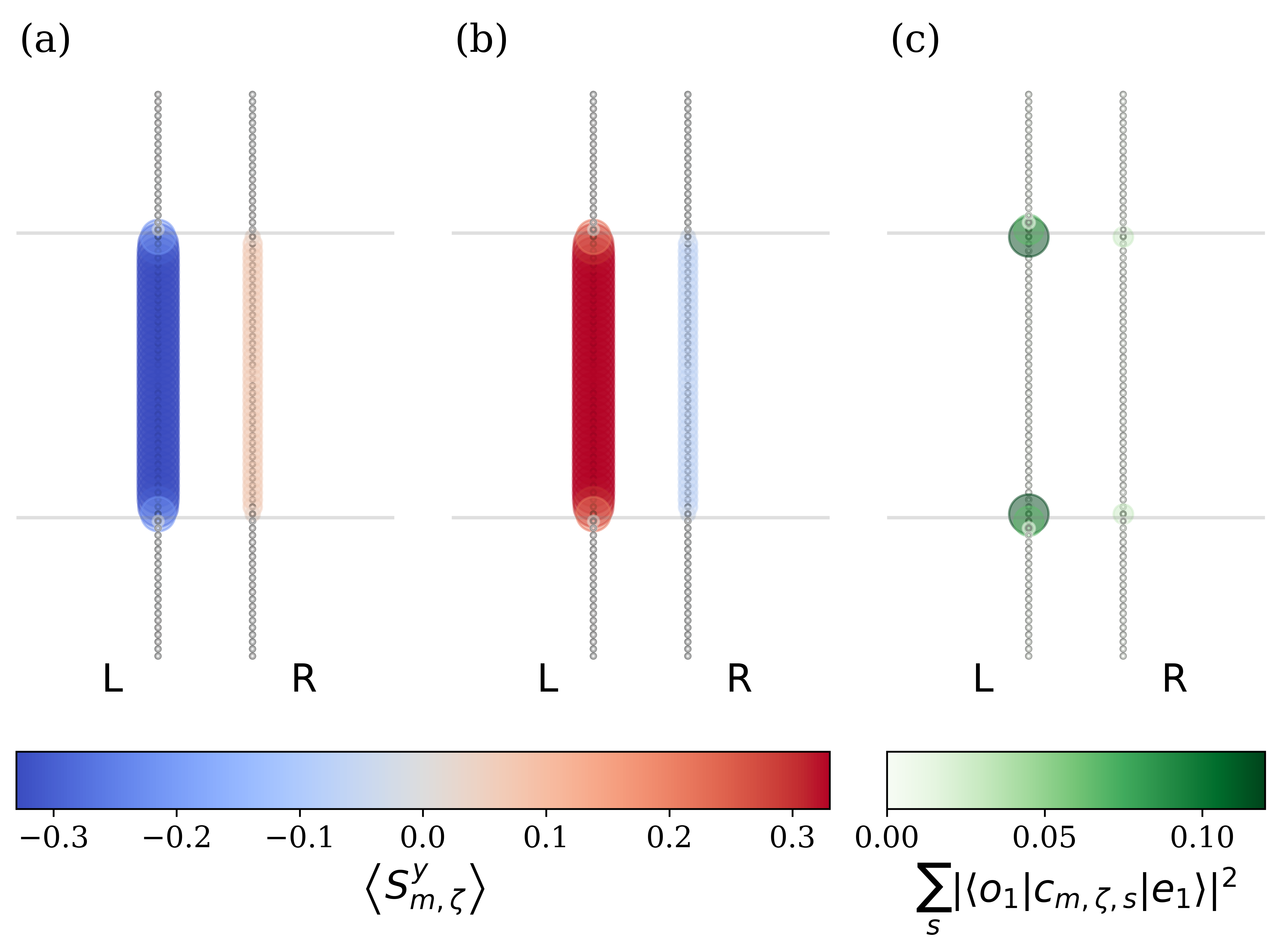}
  \caption{Different local quantities evaluated in the fourfold degenerate subspace with $V/t=2.2$ and $\Delta'=\mu=0$ for a setup depicted in 
Fig.~\ref{fig:sketch_kinetic} (b) with $N_1=N_3=20$ and $N_2=40$. (a) and (b) depicts the expectation value of the local spin momentum $\langle e_1 | S^y_{m,\zeta} | e_1 \rangle$ and $\langle e_2 | S^y_{m,\zeta} | e_2 \rangle$ for the even subspace. Up to numerical precision, states in the odd subspace, $|o_1\rangle$ and $|o_2\rangle$, show the same pattern. (c) depicts the matrix element $ \sum_{s}\left|\langle o_1 | c_{m,\zeta, s} | e_1 \rangle\right|^2$ of the local annihilation operator $c_{m,\zeta,s}$ between states of the even and odd subspace with the same magnetization pattern.}
  \label{fig:local}
\end{figure}

\section{Identification of parafermions}
\label{sect:para}
For the remainder of the work, we focus on the thorough analysis of the phase with fourfold degeneracy to reveal its genuine parafermionic nature.
Parafermionic zero modes have to satisfy the following checklist.
First, the phase hosting these exotic zero energy modes should be characterized by zero energy localized single particle excitations \cite{Kitaev_2001}. Second, the ground state manifold should be robust against external perturbations\cite{alicea_quantum_comp}. In our model, time-reversal symmetry preserving perturbations, \emph{i.e.} external electrostatic fields, are expected to preserve the ground state degeneracy. Perturbations breaking time-reversal are expected to split the zero energy modes \cite{zhang,orth2015non}.
Finally, to distinguish the parafermionic phase from time-reversal symmetric Majorana modes a characteristic Josephson signal will serve as a definitive fingerprint. In the case of Majorana fermions a $4\pi$ periodic Josephson effect is expected \cite{TRITOPS_JOSEPHSON_PhysRevB.90.020501,TRITOPS_JOSEPHSON_PhysRevLett.111.056402}, while $\mathbb{Z}_4$ parafermions exhibit $8\pi$ periodicity\cite{zhang,orth2015non,Peng2016,Vinkler2017,alicea_fendley} .

In this section first, we show that the zero-energy modes are characterized by edge-localized single-particle excitations, robust against electrostatic disorder, as one expects from the bulk-boundary correspondence.
Second, we observe a $8\pi$ periodicity in the Josephson current providing the definitive signature of the $\mathbb{Z}_4$ parafermionic zero modes~\cite{zhang}.

\subsection{Local characteristics}
\label{sect:loc}
The fourfold degeneracy of the ground state is resilient against fluctuations of the electrostatic potential. In our calculations, we find that the expectation value of the local electron density 
\begin{equation}
n_{m,\zeta}=c^\dagger_{m,\zeta,\uparrow}c_{m,\zeta,\uparrow}+c^\dagger_{m,\zeta,\downarrow}c_{m,\zeta,\downarrow},    
\label{eq:density}
\end{equation}
is identical for all four states while the off-diagonal matrix elements of $n_{m,\zeta}$ are numerically negligible. Accordingly, no local electrostatic potential configuration exists which can split the four-fold degeneracy of the ground state.

On the contrary, breaking time-reversal symmetry with a local magnetic field lifts the degeneracy of the ground state owing to the emergence of uneven diagonal matrix elements in the perturbative Zeeman term. 
Focusing on the states of the even subspace $|e_1\rangle$ and $|e_2\rangle$, we observe that the expectation value of $S^y_{m\zeta}$, localized to the region with interactions, differs in sign as depicted in Fig.~\ref{fig:local} (a) and (b). While similar behaviour is found in the odd subspace, all the other matrix elements related to the local spin operators are proved to be negligible.  Note that the magnetization pattern exhibited by the ground state manifold is a clear consequence of the considered interaction \eqref{eq:interaction-spin} favouring an anisotropic spin configuration in the $y$ direction. 

Zero-energy single-particle excitations, in the ground state manifold, are exponentially localized to the interfaces of the superconducting and interacting regions, as depicted in Fig.~\ref{fig:local} (c). This is again an indicator of two topological zero modes attached to the boundary of the interacting region\cite{Kitaev_2001}. However to clearly distinguish these zero-energy excitations from time-reversal invariant pair of Majorana bound states\cite{TRITOPS_PhysRevB.89.220504} further analysis is necessary. 

\subsection{Josephson spectrum}
\label{sect:josephson}
\begin{figure}[!h]
  \includegraphics[width=0.45\textwidth]{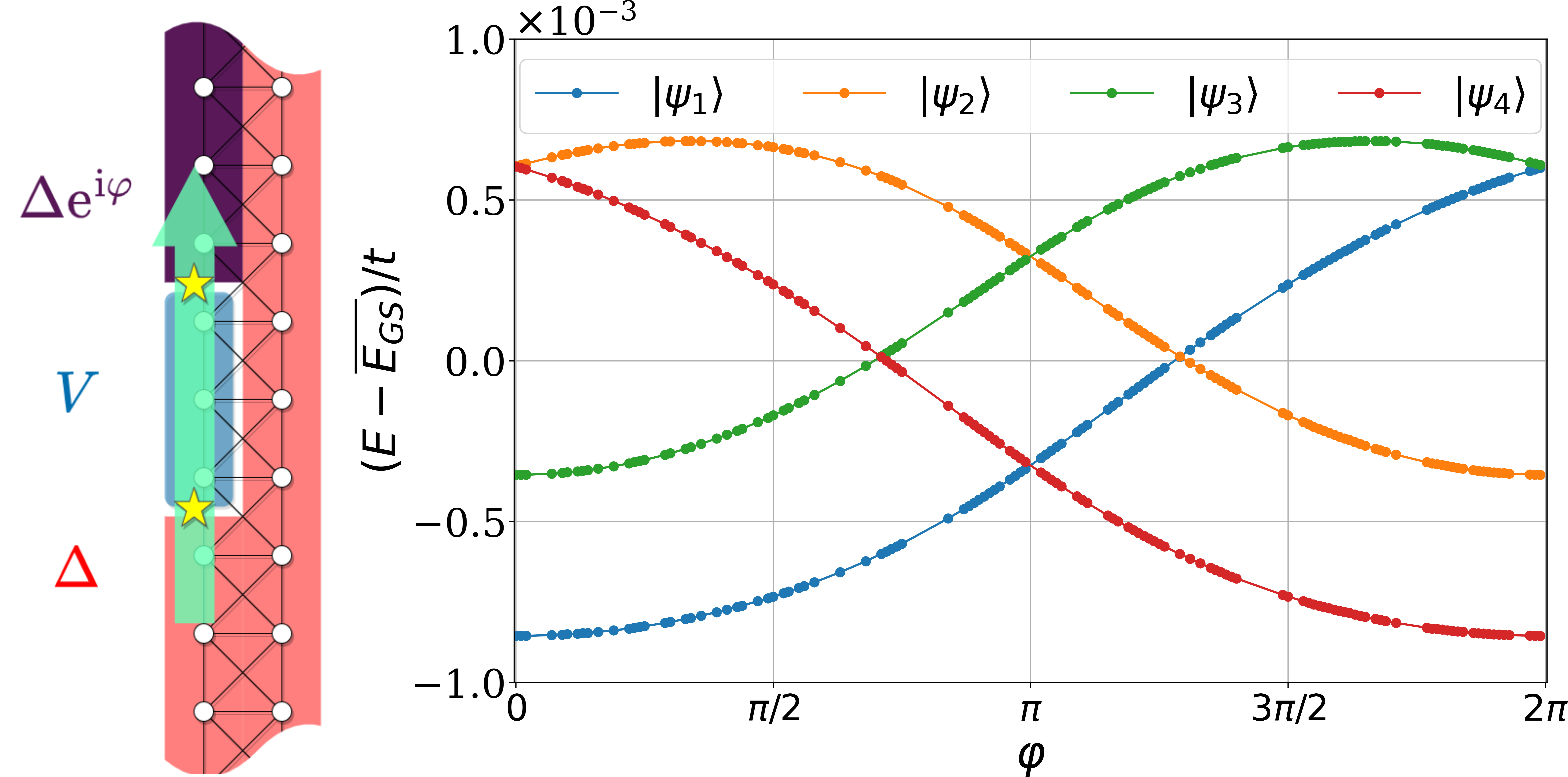}
  \caption{Schematic representation of a phase bias induced Josephson current crossing two parafermionic zero modes (left panel) and Josephson spectrum (right panel) of a junction with $N=8$ sites and interaction strength $V/t=2.2$.}
  \label{fig:josephson}
\end{figure}

The Josephson spectrum, depicted in Fig.~\ref{fig:josephson} decisively reveals the presence of parafermionic zero modes in the considered model. 
The evolution of the energy of the localized four modes as the function of phase bias $\varphi$ in between two superconducting terminals can be used to characterize anyonic excitations\cite{4pi_PhysRevB.79.161408}. In particular, time-reversal invariant Majorana modes show a $4\pi$ periodic modulation\cite{TRITOPS_JOSEPHSON_PhysRevB.90.020501,TRITOPS_JOSEPHSON_PhysRevLett.111.056402}, while $\mathbb{Z}_4$ parafermions exhibit $8\pi$ periodicity~\cite{alicea_fendley}.

The introduced model allows for the isolation of the Josephson current passing through two parafermionic excitations localized at the boundary of the interacting region.
In Fig.~\ref{fig:josephson} we consider a short junction of $N=8$ sites, thus the degeneracy is slightly lifted. At $\varphi=0$ time-reversal symmetry protects the twofold degeneracy of states with odd parity while no such restrictions apply for even states. 
Initializing the system in $|\psi_1\rangle$ we can continuously follow its progression as the phase bias $\varphi$ is tuned. At $\varphi=\pi$ as the consequence of the joint manifestation of the parity anomaly and time-reversal symmetry, the spectrum exhibits two distinct degeneracies \cite{Vinkler2017}. Continuing further, a crossing between $\varphi=\pi$ and $\varphi=3\pi/2$ is protected by local parity of the junction. Arriving at $\varphi=2\pi$ the system evolves smoothly in $|\psi_2\rangle$. We need a further three cycles, that is a total of $8\pi$ shift of the phase bias to recover the initial state. In Apps.~\ref{sect:App_reorthogonalization} and \ref{sect:App_Josephson}, we give additional details regarding the procedures for obtaining the presented Josephson spectrum.

\section{Minimizing the required anisotropy by isotropic exchange}

\label{sect:Heisenberg}
As we have shown above, the stability of the parafermionic phase requires a considerable anisotropic exchange interaction. 
Below, we give numerical evidence that the parafermionic phase can be stabilized for considerably smaller anisotropic exchange interaction as well if a sufficiently large ferromagnetic isotropic Heisenberg exchange interaction is present in the system. 
In this section, we replace the interaction term with the expression:
\begin{equation}
    H_\text{int}=\sum_{m,\zeta}J_{m,\zeta} \mathbf{S}_{m,\zeta}\cdot\mathbf{S}_{m+1,\zeta}+A_{m,\zeta} S^x_{m,\zeta}S^x_{m+1,\zeta},
\end{equation}
where $J_{m,\zeta}$ is the strength of isotropic Heisenberg exchange, while $A_{m,\zeta}$ is single axis anisotropy with $\mathbf{S}_{m,\zeta}=(S^x_{m,\zeta},S^y_{m,\zeta},S^z_{m,\zeta})$ being the vector of electron spin. 

\begin{figure}[!t]
\centering
\includegraphics[width=0.45\textwidth]{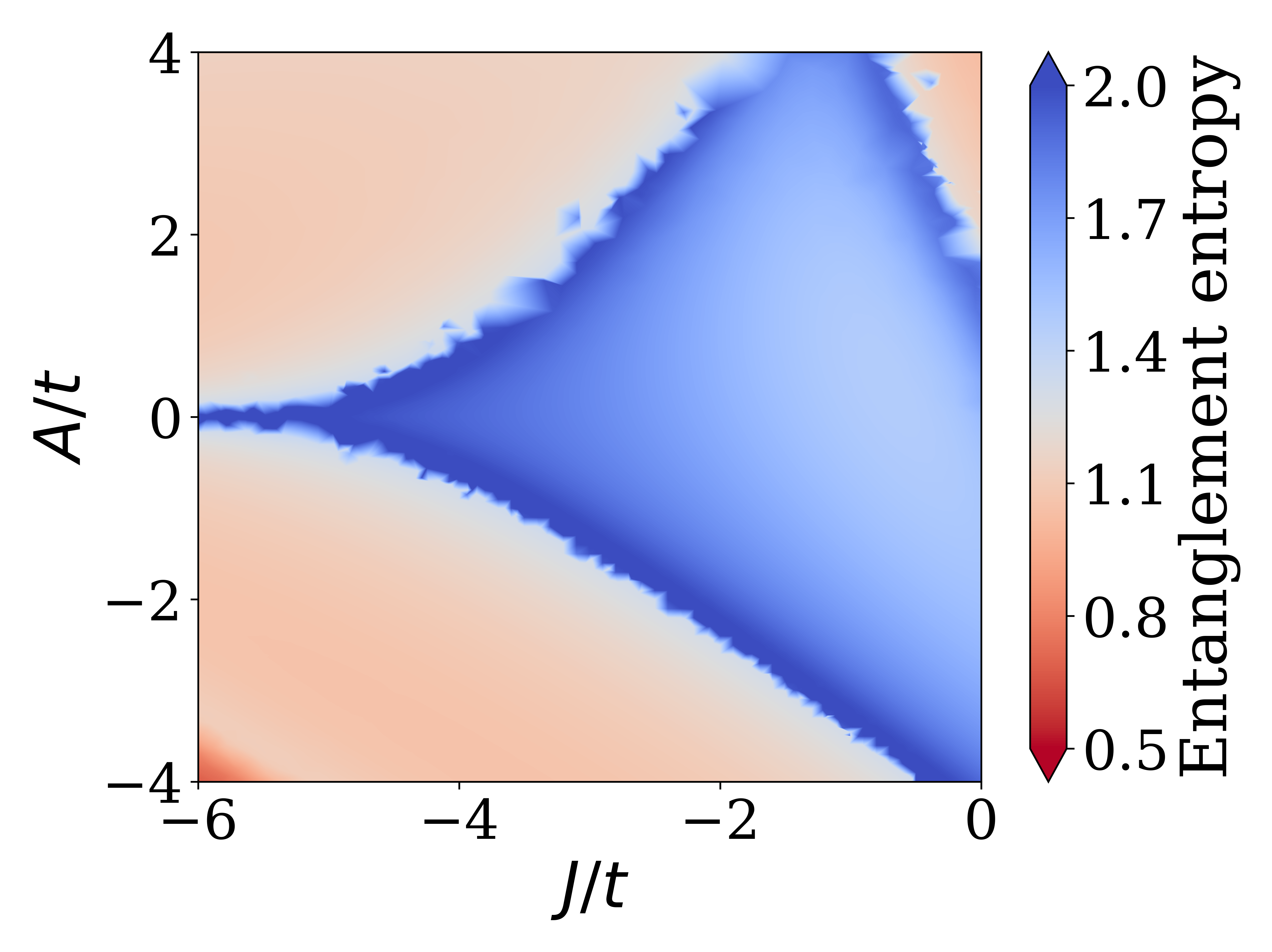}
\caption{Phase diagram for finite isotropic $J$ and anisotropic $A$ exchange, with $N_1=N_3=20$, $N_2=100$. in a geometry similar to the one depicted in Fig.~\ref{fig:sketch_kinetic} (b).}
  \label{fig:J_A_phase}
\end{figure}

\begin{figure}[!t]
\centering
\includegraphics[width=0.45\textwidth]{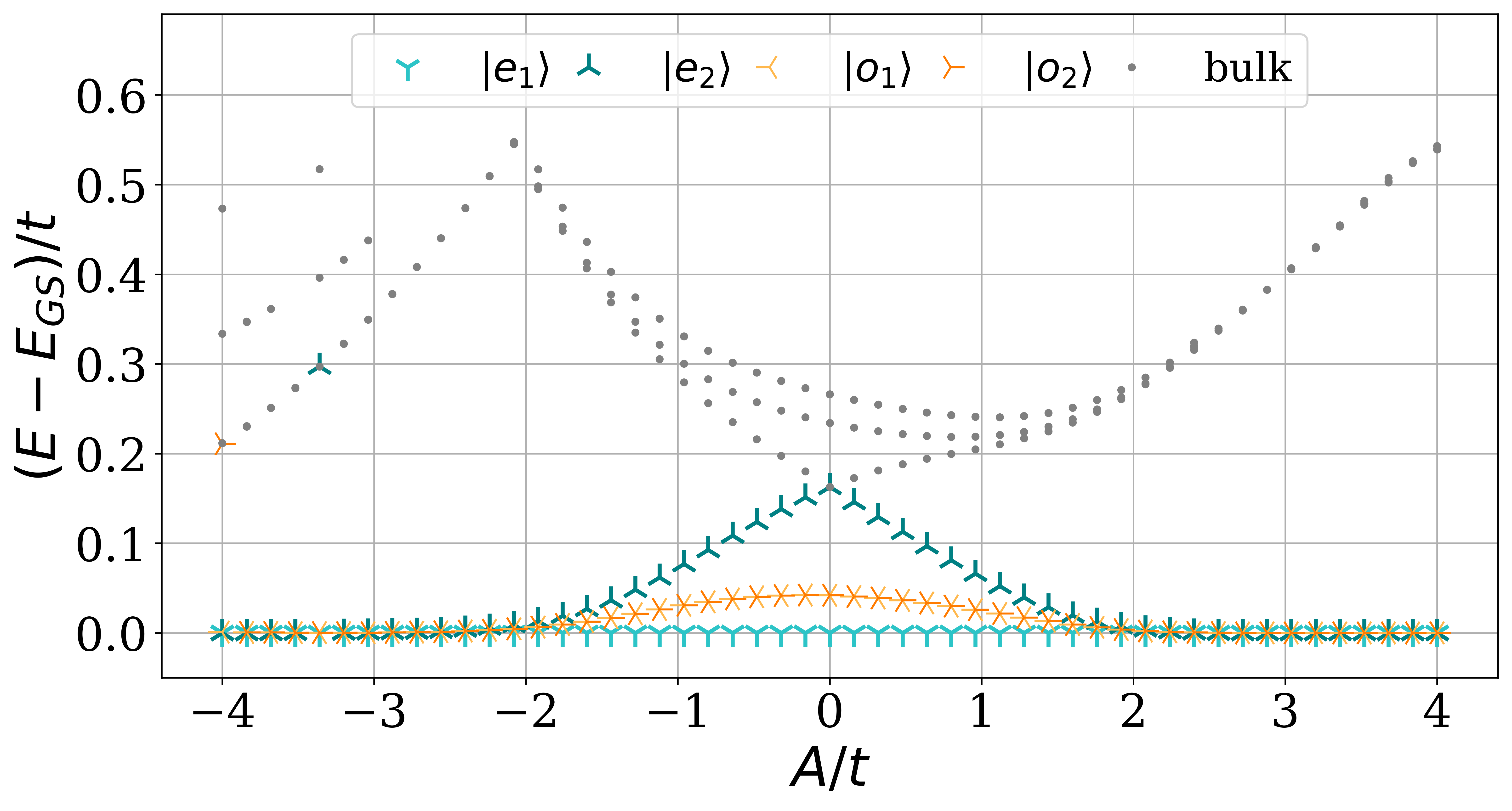}
\caption{Excitation spectrum as the function of the anisotropic exchange $A$ for $J=-4t$.}
  \label{fig:J_A_spectrum}
\end{figure}

The phase diagram of the model is depicted in Fig. \ref{fig:J_A_phase}. 
In the phase diagram, for sufficiently large negative $J$ two phases with reduced entanglement entropy can be discerned which are separated by a critical line.
The excitation spectrum evaluated at $J/t=-4$ as depicted in Fig. \ref{fig:J_A_spectrum} shows that both regions are characterized by a fourfold degenerate ground state and have a considerable excitation gap. 
At around zero anisotropy the two phases are separated by a metallic critical region. 
Further numerical investigation shows that the two low-entropy regions are characterized by a parafermionic $8\pi$ periodic Josephson spectrum.

\section{Discussion}
\label{sect:discuss}
We constructed a lattice model with explicit time-reversal symmetry which is capable of describing the edge states of a topological insulator in the presence of superconductivity and interaction. 
We identified the different phases in the model, in particular, we found a parameter regime where the model hosts parafermionic excitations. 
Our model highlights new realization avenues for these exotic excitations in quantum dot arrays and implies a bottom-up approach for germinating parafermionic zero modes.

Ladder-like geometries consisting of four\cite{FOUR_QDOT_LADDER_2021} and more recently eight quantum dots\cite{EIGHT_QDOT_LADDER_hsiao2023exciton} have been fabricated where hopping, spin-orbit coupling and interactions could be controlled with high accuracy. Quantum dot arrays in conjunction with topological superconductivity have been studied theoretically\cite{SC_DOT_TUNABILITY_THEORY_PhysRevLett.129.267701} and realized experimentally\cite{TWO_DOT_KITAEV_EXP_2023realization,THREE_DOT_KITAEV_EXP_bordin2023crossed}.
Crucially in these systems the relevant energy scales for superconducting pair potential, hopping, and spin-orbit coupling can be tuned and are roughly in the same typical energy scale of 10-100 meV, thus providing a versatile platform for implementing the proposed model. 
Ultracold atomic ladders \cite{CA_Ladder_li2013topological} in combination with spin-orbit coupling mimicking mechanisms\cite{CA_SOC_SPIELMAN1_lin2011spin,CA_SOC2_galitski2013spin,CA_SOC3_SPIELMAN2_valdes2021topological} could also provide an alternative route for the realization of our model.
In line with frameworks employing the QSH phase, layered van der Waals materials may provide a fresh realization pathway as all necessary ingredients can be found in these systems. The QSH effect has been observed in WSe$_2$\cite{vdW_TI_WSe_ugeda2018observation}, superconductivity was measured in twisted bilayer graphene\cite{twisted_bilayer_superc_cao2018unconventional} and 2H-NbS$_2$ samples\cite{vdW_superconductor_devarakonda2020clean}. Anisotropic exchange interaction, crucial for our model, can be also realized using magnetic van der Waals materials~\cite{vdW_anisotropic_exchange_telford2022designing}.

\section{Acknowledgements}
Fruitful discussions with András Pályi, János Asbóth, Yuval Oreg,  Jelena Klinovaja, Daniel Loss, Elsa Prada, Miles Stoudenmire, Gergely Markó, and Alberto Cortijo are greatly acknowledged.
This work was supported by the Ministry of Innovation and Technology and the National Research, Development and Innovation Office of Hungary (NKFIH) within the National Quantum Technology Program (Grant No. 2022-2.1.1-
NL-2022-00004) and project K-115608, K142179. We thank the Frontline – Research Excellence Programme of the NKFIH, Grant No.
KKP133827.
L.O. acknowledges the support of the Bolyai Research Scholarship of MTA and the Bolyai Plusz Scholarship of \'UNKP.
\"O.L. has been supported by NKFIH through Grant Nos.~K134983 and TKP2021-NVA-04,
by the Hans Fischer Senior Fellowship program funded by the Technical
University of Munich - Institute for Advanced Study, and by the Center
for Scalable and Predictive methods
for Excitation and Correlated phenomena (SPEC), funded as part of the
Computational Chemical Sciences Program by the U.S. Department of
Energy(DOE), Office of Science, Office of Basic Energy Sciences, Division
of Chemical Sciences, Geosciences, and Biosciences at Pacific Northwest
National Laboratory.
We acknowledge KIF\"U for awarding us access to computational resources based in Hungary.

\appendix
\label{sect:appendix}
\section{Density matrix renormalization group calculations}
\label{sect:App_DMRG}
We applied the density matrix renormalization group (DMRG) method which is a prominent numerical eigensolver for low-dimensional interacting quantum systems~\cite{Schollwock2005,Verstraete2023}.
The calculations were performed by both the Budapest-DMRG code~\cite{budapest_qcdmrg} and by the ITensor implementation~\cite{itensor} also allowing us to cross-check numerical results.

The accuracy of the DMRG calculations was controlled by the dynamic block-state selection scheme~\cite{Legeza-2003a} where the actual number of retained block states depends on the spectral properties of the corresponding reduced density matrix. 
In the calculations, keeping the truncation error between $10^{-7}-10^{-9}$, the maximal number of block states reached during sweeps was between $M_{\rm max}=1000-2500$. 

The bond dimension of the initial random matrices and the minimum link dimension were set to the same value between $30-100$.
In the DMRG chain representation, the sites of the two legs were arranged in alternating order in order to minimize the distance between correlated sites.
Excited states were obtained in an iterative manner by adding a projector, constructed from the previously computed states to the Hamiltonian with a penalty factor of magnitude $20t-100t$.
While one can readily show the time-reversal symmetry of our proposed model, considering a system with explicit superconductivity,  particle parity remains the only conserved Abelian quantum number that DMRG calculations can make use of.
Even though the limited number of quantum numbers foreshadows a challenge for the accurate numerical treatment of the problem, we found that the substantial gap induced by the different gap generation mechanisms in the system makes the DMRG simulations rather manageable even for system sizes of a couple of hundred sites.
In fact, in the DMRG calculations, we treated ladder models with up to 220 rungs to study finite-size effects. 

\section{High-precision solution---post-DMRG analysis}
\label{sect:App_reorthogonalization}
The quality of the many-body states $\{|\psi_i\rangle\}_{i=1}^{n}$ delivered by the DMRG algorithm with energy $\{\epsilon_i\}_{i=1}^{n}$ was also monitored by the variance~\cite{Hubig2018} $v_i =\langle \psi_i| H^2 |\psi_i\rangle-\epsilon_i^2$, which measures the non-eigenstate content of the computed DMRG  states, was found to be in the order $v/t^2 \approx 10^{-3}$ in all of our calculations. We note that this translates to $\approx 0.03t$ accuracy in energy. 
Since the key objective of the present work was to understand a phase with a highly degenerate ground state, the quality of the predictions in the ground state manifold had to be drastically increased. In particular, for the resolution of the delicate details of the Josephson spectrum an accuracy beyond $10^{-4}t$ was required for the relevant system sizes. 
Instead of performing further DMRG calculations with more stringent precision criteria, and hence demanding substantial computational resources, we performed the following orthogonalization procedure. 
Considering the set of $\{|\psi_i\rangle\}_{i=1}^{n}$ many-body states obtained by DMRG as basis states we re-expanded the Hamiltonian of the model.
Thus, introducing the effective Hamiltonian and overlap matrix $h_{i, j} = \langle \psi_i | H | \psi_j \rangle$, and $s_{i, j} = \langle \psi_i | \psi_j \rangle$ respectively, we solved the generalized eigenvalue problem $h a_p = E_p s a_p$.
Note that the resulting matrix equation of low rank, which equals the number of roots kept in the DMRG calculations, is readily solved by standard means. The obtained $E_p$ eigenvalues give appropriate accuracy for the true spectrum of the system. Using the $a_p$ coefficient vectors the corresponding wavefunctions are obtained as $| \Psi_p \rangle = \sum_i (a_p)_i | \psi_i \rangle$.

\section{Finite scaling of the gaps}
\label{sect:App_spectrum_scaling}
\begin{figure}
\centering
  \includegraphics[width=0.45\textwidth]{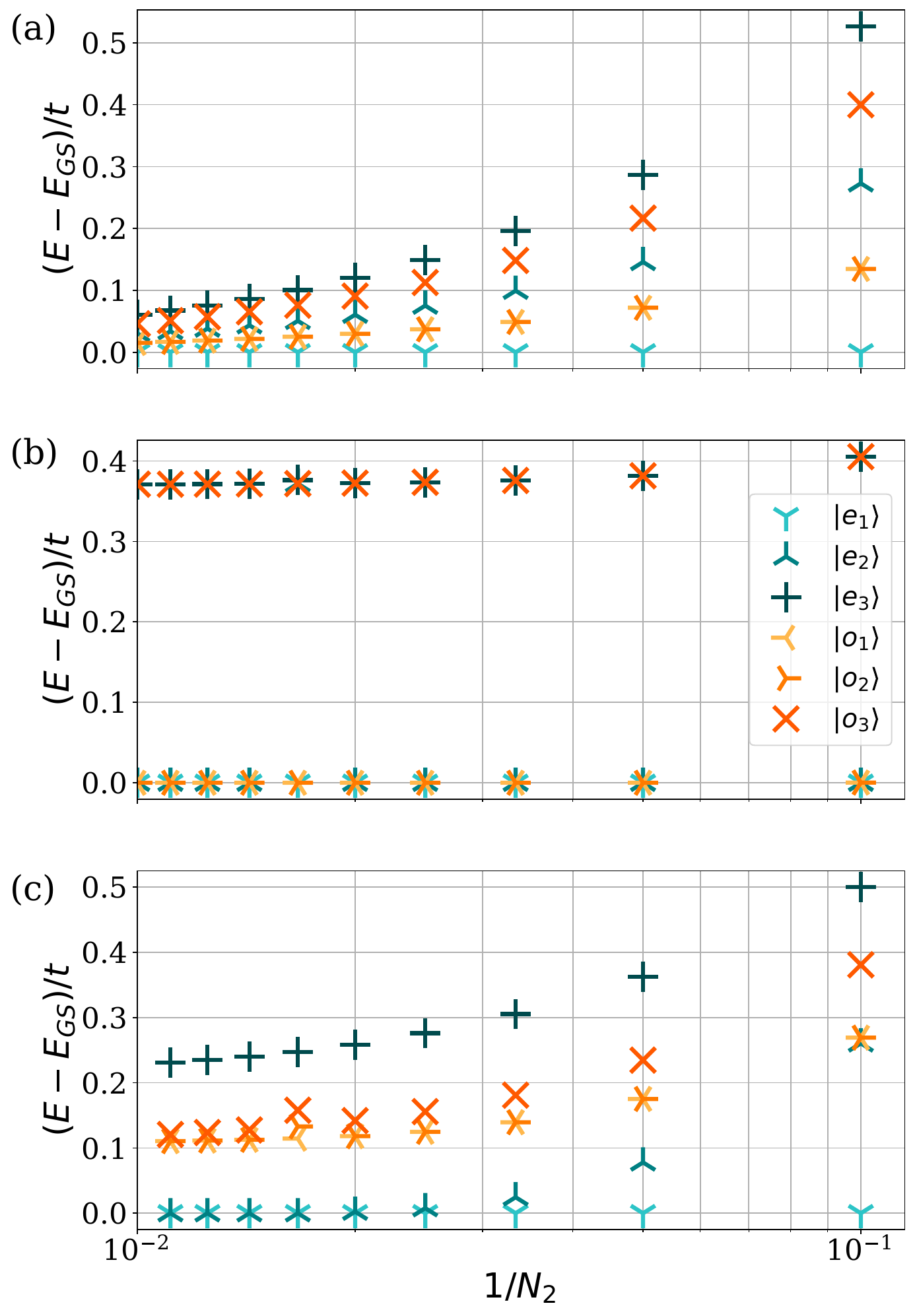}
  \caption{Excitation energies with respect to the inverse of the system size for interaction strength $V/t=$ 0.2, 2.1, and 3.5 in subplot (a), (b) and (c) respectively. Note that the used model setup is sketched in Fig. 1 (b). In the calculations, we set $N_1=N_3=20$, and  $N_2$ was varied. 
  }
  \label{fig:lengthscales}
\end{figure}

The energy of the low-lying excitations with respect to the system size is presented in Fig.~\ref{fig:lengthscales} in a representative point of each phase which reveals the strikingly different structure of the excitations.
For the metallic phase observed for weak interaction strength, as shown in Fig.~\ref{fig:lengthscales} (a), the gaps are vanishing for increasing system size.
On the contrary, for strong enough interactions, see Fig.~\ref{fig:lengthscales} (b) and (c), the excitation energies converge to a finite value in the thermodynamic limit.
Most notable for intermediate couplings illustrated in Fig.~\ref{fig:lengthscales} (b), the ground state is found to be fourfold degenerate for arbitrary system size implying already the potential emergence of parafermionic excitations.

\section{Detailed analysis of the Josephson spectrum}
\label{sect:App_Josephson}
\begin{figure*}[h]
\includegraphics[width=\textwidth]{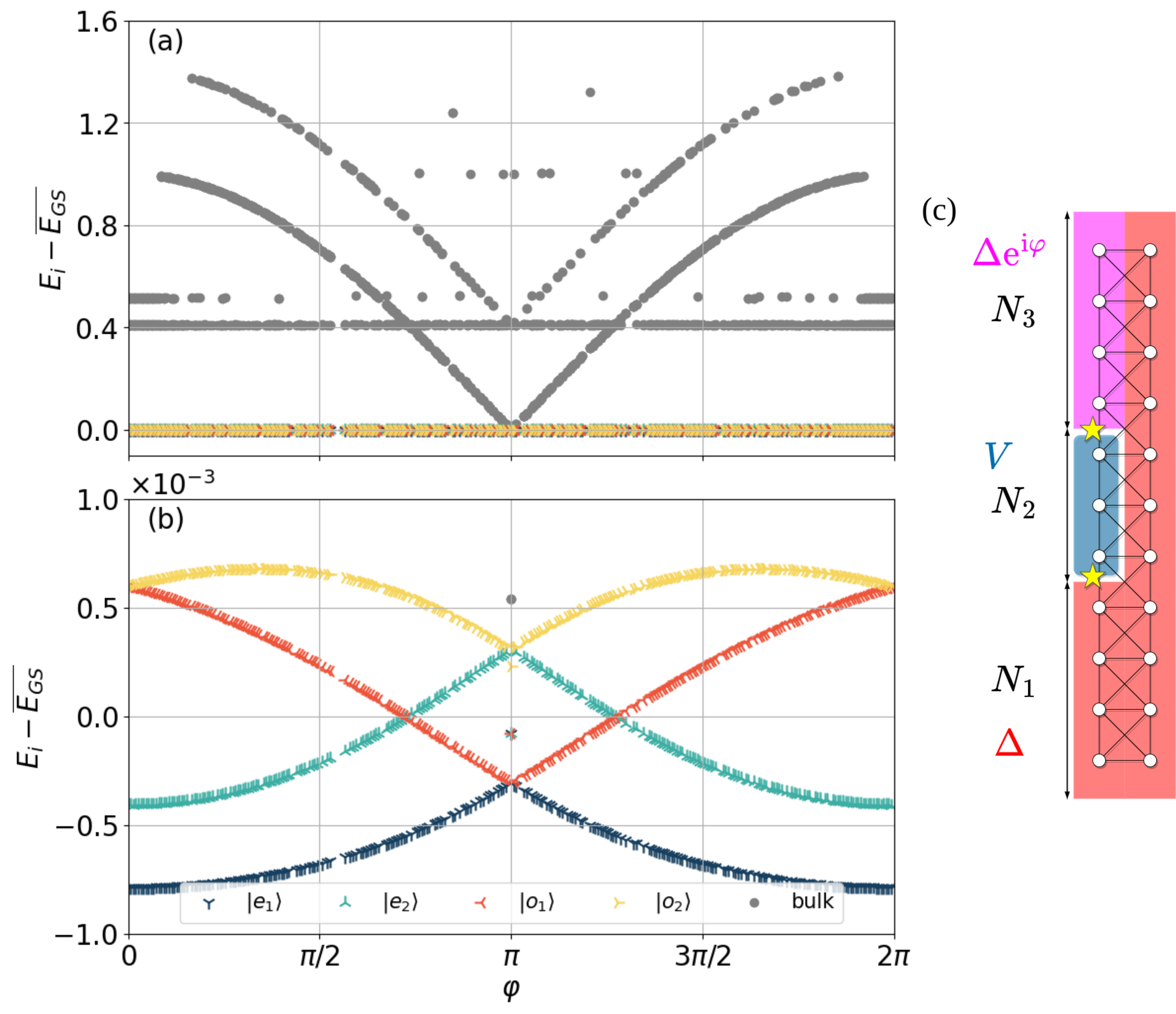}
\caption{Excitation spectrum (a), zoomed to the ground state manifold (b) for a setup depicted in (c) with system size $N_1=20,\,N_2=8,\,N_3=20$, interaction strength $V/t=2.2$ and pair potential $\Delta/t=1.0$ as the function of phase bias $\varphi$. The excitation spectrum is calculated with respect to the mean energy $\bar{E_GS}$ of the four lowest eigenstates $|e_{1/2}\rangle,|o_{1/2}\rangle$. }
  \label{fig:simple_JJ_details_1}
\end{figure*}

In this section, we provide details regarding the calculation of the Josephson spectrum. We consider two setups. In the first configuration, the system has two interfaces binding parafermionic zero modes resulting in a fourfold degenerate ground state manifold. In the second we consider four interfaces, two pinning parafermions and two harbouring Majoranas. If system sizes are appropriately chosen the Josephson current passing through the two parafermions can be studied.
The two setups can be used to understand the nature of band crossings in the Josephson spectrum.

\subsection*{Two interfaces}
Let us focus first on the configuration depicted in Fig.~\ref{fig:simple_JJ_details_1}.
In this setting, a relatively short $N_2=8$ junction induces a small $~10^{-3}t$ splitting in the fourfold degenerate ground state.

As fermionic parity is a good quantum number in the whole model, we can discuss even and odd states separately. At $\varphi=0$ the degeneracy of the even parity of states is lifted while the degeneracy of the odd states remains guaranteed by time-reversal symmetry. As $\varphi$ is changed a crossing of $|e_2\rangle$ and $|o_1\rangle$ can be observed. This crossing is strictly protected by fermion parity. 
At $\varphi=\pi$ in Fig.~\ref{fig:simple_JJ_details_1} (b) further two crossings can be observed accompanied by a seemingly discontinuous shift of the parity of states. This crossing is again protected by time-reversal symmetry. Zooming out in subfigure (a) we observe a bulk state plunging to zero energy at exactly $\varphi=0$. Inspecting the local properties of this state reveals that it is localized at the edge of the system, far from the interface. The appearance of this state at low energies, also visible in (b) as a single grey point at $\varphi=0$, explains the observed discontinuous evolution of the parity eigenvalues as this state carries charge from the interaction region. This effect is often labelled as the parity anomaly\cite{Vinkler2017}.
As $\varphi$ is tuned further we cross another degeneracy protected by parity. Finally, at $\varphi=2\pi$, all states seem to return to their original position.
At this point, we are faced with a dilemma. If we prepare the state say $|e_1\rangle$ How will it evolve as we tune across the seeming parity discontinuity at $\varphi=\pi$. To address this issue we shall focus on a slightly extended system where there are no high-energy states interfering with the ground state.

\subsection*{Four interfaces}

\begin{figure}
\includegraphics[width=0.15\textwidth]{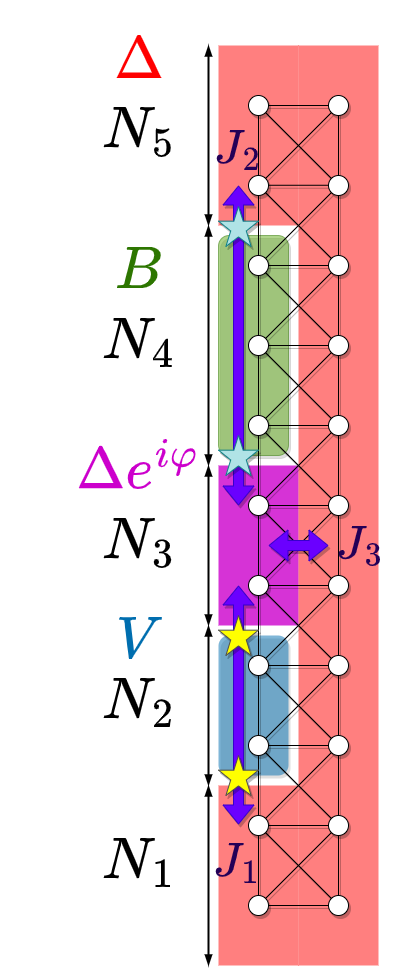}
\caption{\label{fig:JJ_config} System configuration used for the calculation of the Josephson spectrum with four interfaces.}
\end{figure}
\begin{figure*}
\includegraphics[width=\textwidth]{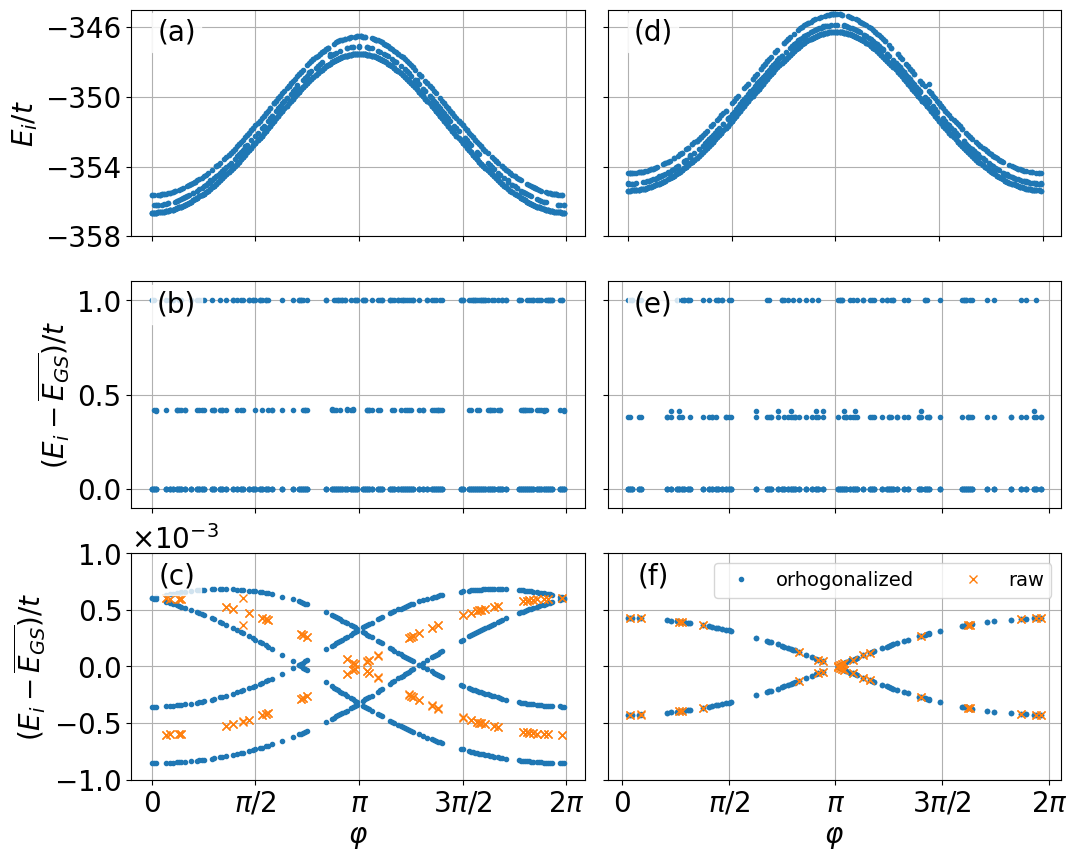}
  \caption{Josephson spectrum for two different geometrical setups. In (a)-(c) $N_2=8$, while all other length scales are $20$, in (d)-(f) we set $N_4=8$ and set the rest again to $20$. In (a) and (d) the evolution of the first couple of many-body states is depicted, while (b) and (e) show excitation spectrum compared to the degenerate ground state (c) and (d) zoom in on the low energy manifold. In all plots $\Delta/t=B/t=1.0$ and $V/t=2.2$. Data in these plots were obtained in the global odd sector. The even sector exhibits the same spectrum with a maximal difference not bigger than $10^{-6}t$. That is considering both parities in (c) each blue point is doubly degenerate while in (f) quadruple degeneracy is observed.}
  \label{fig:JJ_details_1}
\end{figure*}

To acquire the Josephson spectrum without the high-energy bulk state crossing the ground state manifold, we considered a system with the geometry depicted in Fig.~\ref{fig:JJ_config}. On top of the kinetic background, represented by the lattice structure in the figure, we consider interfaces between two superconducting regions, denoted by reddish and magenta colours in the figure. On the left side of the system, we consider an interacting region, marked by a bluish colour and characterized by interaction strength $V$, and a region with a finite magnetic field pointing in the $x$ direction with magnitude $B$. Each region has a length labelled by $N_i$. We need to resort to this geometry To avoid interface states between the superconducting regions, localized far from the interacting region, which traverses the excitation gap as $\varphi$ is varied, thus masking crucial parts of the spectrum. Introducing a region with a magnetic field proved to be a particularly useful approach. This choice pins the aforementioned interface states to zero in the form of two Majorana fermions, marked by light blue stars in the figure. We can identify three Josephson junctions in the considered geometry, indicated by blue arrows in the figure. The magnitude of the Josephson current for the three junctions is denoted by $J_i$. Changing the $N_i$ geometrical parameters the separate $J_i$ contributions to the current behave differently. $J_1$, the contribution through the interaction region is exponentially suppressed as $N_2$ is increased, similarly $J_2$ is diminished if $N_4$ is enlarged. $J_3$, on the other hand, depends linearly on $N_3$. 
Thus typically we expect to have a large modulation due to $J_3$ with period $2\pi$. Increasing $N_4$ to the limit where the two Majorana fermions decouple we can concentrate on the signal coming from tunnelling through the localized zero energy excitations marked by yellow stars at the edge of the interacting region.
We show the results of such a calculation in Fig.~\ref{fig:JJ_details_1} (a)-(c). The large-scale $2\pi$ periodic modulation is evident from the many-body spectrum in (a). However, if we focus on the low energy excitations shown in (c) then the orthogonalized excitation spectrum shows a characteristic $8\pi$ modulation! 

\begin{figure}
\centering
  \includegraphics[width=0.45\textwidth]{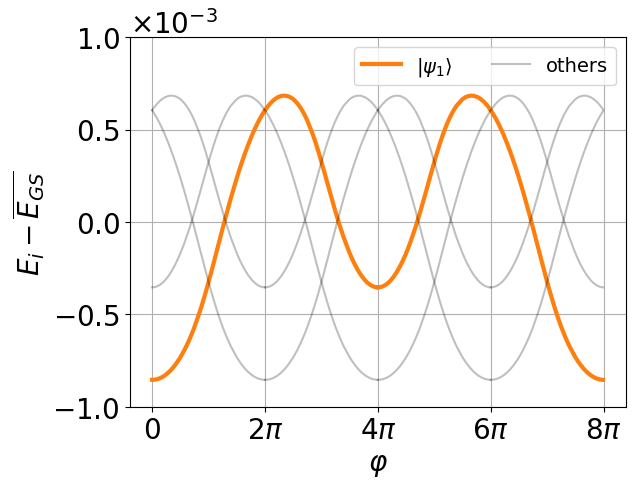}
  \caption{Continuous Josephson spectrum obtained for the globally odd states through linear sum assignment from the overlap matrices of orthogonalized states. All parameters are the same as in Fig.~\ref{fig:JJ_details_1} (c). }
  \label{fig:JJ_details_2}
\end{figure} 

This observation confirms that these excitations are indeed parafermionic. 
We have to make an important observation regarding the raw DMRG data. The raw data exhibit a $4\pi$ periodic modulation compared to the orthogonalized values. Thus one has to be careful when drawing conclusions based on these values alone. Of course, the comparatively poor accuracy of the raw results is at fault. Given enough computational resources the raw data can be made more precise, however, orthogonalization is much more practical for this use case.

We present the results obtained for the opposite case in Fig.~\ref{fig:JJ_details_1} (d)-(f). In this setup, the two parafermionic modes are the ones that decouple due to an increased separation, and the Majorana fermions remain hybridized due to a shortened junction. Thus we expect $2\pi$ modulation for the large-scale structure and upon close inspection a $4\pi$ periodic Josephson signal when we concentrate on the ground state manifold, as it is evident from the figure our expectations are fulfilled. Raw DMRG spectra in this case are qualitatively in agreement with the orthogonalized values.
We note that we only show data at $\varphi$ values where the raw DMRG calculations or the spectra obtained after orthogonalization possess a fourfold degenerate ground state with a tolerance higher than $10^{-2}t$. As in the applied DMRG implementation, excited states are found one after the other, in practice nothing guarantees that at a given value of $\varphi$, all states in the degenerate ground state manifold will be found. As the starting point of the DMRG calculations is a random state, this numerical issue can be solved by redoing the calculation at a given $\varphi$ appropriately many times. 
Further processing the obtained data shown in Fig.~\ref{fig:JJ_details_1} (c) we can make the $8\pi$ periodicity of the Josephson current through the interacting region more explicit. 
By considering the overlaps $\langle \Psi_i(\varphi) |\Psi_j(\varphi+\delta) \rangle$ between orthogonalized states at neighboring $\varphi$ and solving the assignment problem\cite{kuhn1955hungarian} one can arrange states according to their evolution in $\varphi$. The result shown in Fig.~\ref{fig:JJ_details_2} demonstrates how the energy of these states develop as we tune $\varphi$. 
To understand this picture it is useful to discuss how the low-energy manifold of the system is comprised in a given parity sector. First, we focus on a qualitative analysis.
The Majorana zero mode localized in the region with a magnetic field can either be filled or empty giving two states with differing local parity which we denote by $|e_{M} \rangle$ and $|o_{M} \rangle$ for even and odd parity respectively.
The region with interactions on the other hand hosts four states. We have two even $|e_{1/2} \rangle$ and two odd $|o_{1/2} \rangle$ parity states localized to this region. 
The global even subspace of the system is spanned by the four states $|e_{1/2} \rangle\otimes|e_{M} \rangle$ and $|o_{1/2} \rangle\otimes|o_{M} \rangle$, while the global odd subspace has the remaining four states given by $|e_{1/2} \rangle\otimes|o_{M} \rangle$ and $|o_{1/2} \rangle\otimes|e_{M} \rangle$. 
Focusing on the globally odd states at $\varphi=0$ depicted in Fig.~\ref{fig:JJ_details_2} the lowest energy state, $| \Psi_1(0) \rangle$, is composed as $|e_{1} \rangle\otimes|o_{M} \rangle$. As we tune $\varphi$ up to $2\pi$ three crossings are observed. The first two are protected due to local parity conservation since they evolved from the $|o_{1/2} \rangle\otimes|e_{M} \rangle$ states while the degeneracy at $2\pi$ is the result of time-reversal symmetry. At this point, an integer charge is transferred across the junction and we smoothly evolved from the initial state to $|o_{2} \rangle\otimes|e_{M} \rangle$ while flipping the parity of the Majorana bound state. As we tune $\varphi$ further towards $4\pi$ we traverse again crossings protected by local parity and cycle another unit of charge across the junction arriving at $|e_{2} \rangle\otimes|o_{M} \rangle$. At $6\pi$ this turns in to $|o_{1} \rangle\otimes|e_{M} \rangle$. Finally at $8\pi$ we return to $|e_{1} \rangle\otimes|o_{M} \rangle$. In the even parity subspace, similar reasoning gives the sequence $|e_{1} \rangle\otimes|e_{M} \rangle \rightarrow |o_{2} \rangle\otimes|o_{M} \rangle \rightarrow |e_{2} \rangle\otimes|e_{M} \rangle \rightarrow |o_{1} \rangle\otimes|o_{M} \rangle 
\rightarrow |e_{1} \rangle\otimes|e_{M} \rangle$.
The observed $8\pi$ periodic Josephson effect serves as the definitive fingerprint for parafermionic excitations. 

\begin{figure*}
\includegraphics[width=\textwidth]{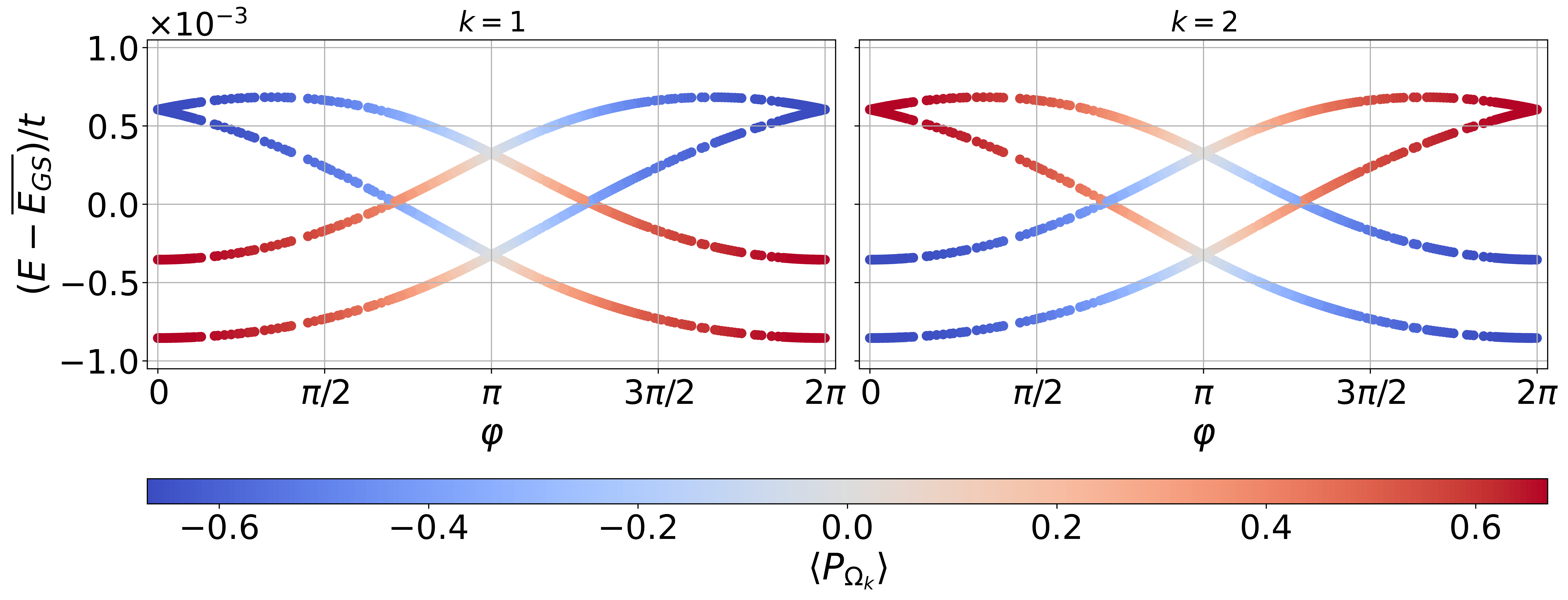}
\caption{Partial parity expectation values for states in the Josephson spectrum in the globally odd sector. }
\label{fig:JJ_parity}
\end{figure*}

Additionally, more quantitative information complementing the discussion above can be gained by evaluating the expectation value of the partial parity operator for a collection of sites $p\in \Omega$: 
\begin{equation}
P_\Omega=\prod_{p\in\Omega,\sigma}\left (-1\right )^{n_{p,\sigma}}.\label{eq:local_parity}
\end{equation}
We cut the system into two parts in the middle of the superconducting region with varying phases (the region with length $N_3$ in Fig. \ref{fig:JJ_config}). That is the first region, which we denote by $\Omega_1$, contains all the sites where interactions are active, thus this is the region with parafermionic zero modes. 
The second region $\Omega_2$ contains the sites where the magnetic field is active, thus this region contains the Majorana zero modes. The results of evaluating the matrix elements of the local parity operator defined in \eqref{eq:local_parity} in the two regions for all states of the globally odd sector is depicted in Fig. \ref{fig:JJ_parity}.

As it can be seen the matrix elements for region $\Omega_1$ for the lowest two eigenstates evolve smoothly from a positive number to a negative number, while the two larger eigenvalues interpolate smoothly from a negative number to a positive. In region $\Omega_2$ a reversed tendency can be observed for all cases! 
For the globally even set of eigenstates the evolution of the local parities in the two regions is identical to the one depicted for region $\Omega_1$ in the globally odd states.
Thus this quantitative analysis is fully in line with the more simple qualitative picture discussed before.

\bibliography{newref}

\end{document}